\documentclass[twocolumn,showpacs,preprintnumbers]{revtex4}
\usepackage{amssymb}
\usepackage{amsmath}
\usepackage{graphicx}
\usepackage{dcolumn}
\usepackage{bm}

\input{tcilatex}
\begin{document}

\title{Nonequilibrium dynamic transition in a kinetic Ising model driven by
both deterministic modulation and correlated stochastic noises}
\author{SHAO Yuanzhi, ZHONG Weirong, HE Zhenhui}
\affiliation{State Key Laboratory of Optoelectronic Materials and Technologies,
Department of Physics, Sun Yat-Sen University, Guangzhou 510275, China}
\email{stssyz@mail.sysu.edu.cn}

\begin{abstract}
We report the nonequilibrium dynamical phase transition (NDPT) appearing in
a kinetic Ising spin system (ISS) subject to the joint application of a
deterministic external field and the stochastic mutually correlated noises
simultaneously. A time-dependent Ginzburg-Landau stochastic differential
equation, including an oscillating modulation and the correlated
multiplicative and additive white noises, was addressed and the numerical
solution to the relevant Fokker-Planck equation was presented on the basis
of an average-period approach of driven field. The correlated white noises
and the deterministic modulation induce a kind of dynamic symmetry-breaking
order, analogous to the stochastic resonance in trend, in the kinetic ISS,
and the reentrant transition has been observed between the dynamic disorder
and order phases when the intensities of multiplicative and additive noises
were changing. The dependencies of a dynamic order parameter Q upon the
intensities of additive noise A and multiplicative noise M, the correlation $%
\lambda $ between two noises, and the amplitude of applied external field h
were investigated quantitatively and visualized vividly. A brief discussion
was given to outline the underlying mechanism of the NDPT in a kinetic ISS
driven by an external force and correlated noises.
\end{abstract}

\keywords{Ising spin system, nonequilibrium dynamical phase transition,
stochastic resonance, correlated noises, TDGL model.}
\pacs{75.10.Hk, 64.60.Ht, 05.10.Gg, 76.20.+q}
\maketitle

\subsection{\textbf{Introduction}}

Ising spin system (ISS) is an ideal theoretical model to describe a variety
of physical phenomena of a uniaxial anisotropy system [1]. The
nonequilibrium dynamic phase transition (NDPT) arises in a kinetic Ising
spin system when a system is driven by both an external field and
temperature simultaneously. Depending upon the combination of temperature
and driving field, there exist in a kinetic ISS two sorts of dynamical
phases, i.e. symmetry-breaking order phase and symmetric disorder
counterpart, and the two sorts of dynamical phases transform into each other
with the variation of driving field and temperature[2\symbol{126}8]. The
time-dependent Ginzburg-Landau (TDGL) equation offers, within an
approximation of mean-field, a more tractable continuum solution to the
kinetic ISS than a discrete approach of Monte Carlo simulation does in
evaluating the NDPT of a kinetic ISS.

Introducing a stochastic fluctuation within a kinetic deterministic system
is a topic of interest and it arouses an extensive concern of
multidisciplinary researchers in recent years. A large quantity of evidence
suggests that the noise plays an important and constructive role in a
nonlinear system [9-17], such as the phase transitions induced by a noise,
the stochastic resonance (SR), the complexity of biology and so on. Among
these, the most challenging subject is associated with the NDPT of a kinetic
ISS driven by a deterministic modulation, e.g. a sinusoidal oscillation, and
a stochastic white noise simultaneously. Unfortunately, the studies on the
NDPT in a kinetic ISS have so far focused simply on a single deterministic
external field [2-7], a deterministic field jointly coupled with either a
sole Gaussian white noise or pulse source [8,18-20]. Few studies have ever
touched on the NDPT of a kinetic ISS including a correlated double noise
source. Zaikin et al. investigated the double SR of bistable system driven
by an external deterministic field with the addition of an uncorrelated
double noise [13]. Denisov et al., however, studied the NDPT induced simply
by a double multiplicative noise at a zero deterministic field [16].
Furthermore, Jia et al. reported the reentrant phenomenon appearing in a
bistable system perturbed by correlated multiplicative noises at a zero
deterministic field [14] as well as the SR due to the collaboration of both
correlated noises and deterministic external field [15]. The above studies,
although revealing from the different aspects some individual features of
the NDPT and the SR of a bistable system in different parameter spaces, have
not yet given a clear account for the NDPT of a kinetic ISS, especially the
feature in dynamic responding when the correlation of a multiplicative and
additive noises, the driving field and the system temperature vary
accordingly.

We have provided an insight into the influence of correlated additive and
multiplicative noises on the growth rate of a tumor using the Logistic model
in a previous paper [21]. Here we present some latest results of our study
on the NDPT of a kinetic ISS. We visualize in a clear and vivid way the NDPT
occurring in the multi-parameter space consisting of the correlation
coefficient ($\lambda $) between additive and multiplicative noises as well
as their intensities (A, M,), and the amplitude of deterministic driving
field ($h_{0}$). This paper is organized with five sections, including the
current introductive section. In sec. 2, we give a brief description of a
kinetic ISS and the basic feature of the NDPT, the stochastic TDGL equation
with a correlated multiplicative and additive noise source as well as the
algorithm for solving numerically the relevant Fokker-Planck equation. The
computational results and the relevant analysis, especially the mechanism
underlying the peculiar reentrant NDPT in a kinetic ISS, are presented in
detail in sec3 and 4, respectively. The characteristics of the NDPT in a
kinetic ISS are summarized in sec.5.

\bigskip

\subsection{\textbf{Nonequilibrium dynamical phase transition and
theoretical model}}

Considering the kinetic Ising model of N interacting spins driven by an
external deterministic field, one can express its Hamiltonian as

\begin{equation}
\hat{H}=-\frac{J}{N}\underset{\{i,j\}}{\sum }\vec{S}_{i}\cdot \vec{S}%
_{j}-h(t)\underset{i}{\sum }\vec{S}_{i}  \tag{1}
\end{equation}

\bigskip

where J is the exchange-coupling constant and spin S = $\pm 1$. Symbol i and
\{i,j\} stand for a single spin-site and a spin-pair of the nearest
neighbor, respectively. The time variation of driving field h(t) is written
in a simple cosinoidal form. The stochastic TDGL equation of Eq. (1) within
the approximation of mean-field is described by

\begin{equation}
\frac{\partial m}{\partial t}=-\Gamma \frac{\delta H(m)}{\delta m}+m\xi
(t)+\rho (t)  \tag{2-1}
\end{equation}

\bigskip

H(m) is the usual m$^{\text{4}}$ Hamiltonian for the instant order-parameter
m(t) and consists of the following terms:

\begin{equation}
H(m)=\int d^{d}r\{\frac{r_{0}}{2}m^{2}+\frac{u_{0}}{4}m^{4}-m\cdot h_{0}\cos
(\omega t)\}  \tag{2-2}
\end{equation}%
\qquad

\bigskip

$\Gamma $ and u$_{\text{0}}$ are two phenomenological coefficients. The
reduced temperature r$_{\text{0}}$ is defined as $\backsim $T-T$_{\text{C0}}$%
;T and T$_{\text{C0}}$ are the temperature of system and static critical
temperature, respectively; h$_{\text{0}}$ and $\omega $ denote the amplitude
and frequency of a deterministic driving field; t is the evolution time of
the system; $\xi $(t) and $\rho $(t) represent the multiplicative and
additive Gaussian noises which satisfy the zero-mean and autocorrelation,
expressed as

\begin{equation}
\langle \xi (t)\rangle =0,\langle \xi (t)\xi (t^{\prime })\rangle =2M\delta
(t-t^{\prime })  \tag{3-1}
\end{equation}

\begin{equation}
\langle \rho (t)\rangle =0,\langle \rho (t)\rho (t^{\prime })\rangle
=2A\delta (t-t^{\prime })  \tag{3-2}
\end{equation}

\begin{equation}
\langle \xi (t)\rho (t^{\prime })\rangle =2\lambda \sqrt{MA}\delta
(t-t^{\prime })  \tag{3-3}
\end{equation}

Here M, A and $\lambda $ in eqs.(3) symbolize the intensities of mutually
correlated multiplicative and additive noises, and the correlation intensity
between them, respectively.

To quantify the process of dynamic phase transition, the dynamic order
parameter Q is defined as

\begin{equation}
Q=\frac{\omega }{2\pi }\doint m(t)dt  \tag{4}
\end{equation}

Actually, dynamic order-parameter Q is the period-average of the instant
order-parameter m over evolution time t. Q=0 and Q$\neq $0 correspond to a
symmetric dynamic disorder phase and a symmetry-breaking dynamic order
phase, respectively. And they were also referred to as the
symmetry-restoring oscillation (SRO) and symmetry-breaking oscillation (SBO)
[8]. The purpose of the current paper is to investigate quantitatively the
dependence of Q parameter of a kinetic ISS upon M, A, $\lambda $ and h$_{%
\text{0 }}$combined.

The relevant Fokker-Planck (F-P) equation of eq.(2) is given as with the
initial and boundary conditions specified

\begin{equation}
\frac{\partial p(m,t)}{\partial t}=-2\frac{\partial \lbrack C(m,t)p(m,t)]}{%
\partial m}+\frac{\partial ^{2}}{\partial m^{2}}[B(m)p(m,t)]  \tag{5-1}
\end{equation}

\begin{eqnarray}
C(m,t) &=&(1+M)m-m^{3}+h_{0}\cos (\omega t),  \TCItag{5-2(1)} \\
B(m) &=&Mm^{2}+2\lambda \sqrt{MA}m+A  \TCItag{5-2(2)}
\end{eqnarray}

\begin{equation}
p(m,0)=\delta (m-0),p(-\infty ,t)=p(\infty ,t)=0  \tag{5-3}
\end{equation}

Apparently, partial differential eq. (5-1), owing to its explicit
time-dependence, is non-autonomous in nature and it is feasible to work out
the numerical solution of F-P eq. (5-1), i.e. probability distribution of m
against time t, p(m,t).

According to the definition of Q parameter in eq. (4) as well as the
numerical solution of p(m,t) above, we can figure out finally the dynamic
order parameter Q of a kinetic ISS using

\begin{equation}
Q_{F-P}=\mid \frac{_{\dint\nolimits_{0}^{\tau }\dint\nolimits_{-\infty
}^{\infty }m(t)p(m,t)dtdm}}{_{\dint\nolimits_{0}^{\tau
}\dint\nolimits_{-\infty }^{\infty }p(m,t)dtdm}}\mid  \tag{6}
\end{equation}

Fig. 1a shows an asymptotically steady and symmetric p(m,t) at a zero
driving field and the zero correlation of two sorts noises. Time-dependent
oscillations of p(m,t) are displayed in Fig.1b (nonzero driving field but
zero correlation of noises ) and 1c (nonzero driving field and nonzero
correlation). The symmetry of p(m,t) is destroyed by the mutually correlated
noises and no asymptotically steady solution is available when a
time-dependent driving field is involved.

\bigskip

\ \ \ 

\bigskip

\subsection{\textbf{Results and analysis}}

Fig. 2 displays the dependence of the reduced Q parameter upon both M and A
under the condition of a different correlation intensity $\lambda $. The
dynamic order parameter Q, due to the cooperation of the driving field with
the correlated noises, attains a distinctive resonant peak within a certain
range of M and A, quite similar to the characteristic of a common stochastic
resonance when the ratio of signal versus noise is evaluated. A larger
correlation intensity leads to a higher Q peak as shown in Fig. 2. The
existence of a peaking Q parameter indicates the enhancement of dynamic
ordering in a kinetic ISS if the deterministic driving force could
synchronize with the stochastic force. Judging from the viewpoint of the
NDPT, the resonance-like dependence of Q parameter is a kind of reentrant
phenomenon that occur usually in a peculiar noise circumstance. Both
additive and multiplicative noises give rise to a reentrant trend of the
dynamic order parameter but, according to Fig. 2, the multiplicative noise
has a more evident effect than the additive one does.

\bigskip \bigskip

Fig. 3 exhibits two four-dimensional panoramic plots of the sliced
contour-lines of dynamic order Q within the parameter space of A, M, $%
\lambda $ and A, M, h$_{\text{0}}$,. It has been displayed vividly that a
correlated noise source induces the enhancement of dynamic ordering as well
as the NDPT of reentrant characteristic. In contrast to noise intensity, the
amplitude of a deterministic driving field simply results in a monotonic
trend of Q parameter.

\bigskip

\subsection{\textbf{Discussion}}

When an interacting collaboratively many-body system, such as the ISS in the
present paper, is driven in time by an external field, the system cannot
respond instantaneously to the perturbation due to the relaxational delay of
the system itself [2], and therefore there come along a series of intriguing
nonequilibrium phenomena, such as dynamic response, dynamic hysteresis and
dynamic phase transition. There exist two competing time scales in a kinetic
spin system, namely the oscillation period of an external driving field $%
\tau _{\omega }=2\pi /\omega $and the intrinsic relaxation time of a spin
system $\tau _{s}$. The latter tends to diverge because of the critical
slowing-down when temperature approaches to a critical point. The NDPT takes
place only when the two competing time scales, $\tau _{\omega }$ and $\tau
_{s}$, match each other someway in quantity [3,4]. The involvement of a
noise source introduces the third competing time scale $\tau _{K}$. As for
an activated Brownian particle jumping between the double potential wells,
the noise-induced hopping between the local equilibrium states with the
Kramers rate $\gamma _{K}$ as given as [9],

\begin{equation}
\gamma _{K}=\gamma _{0}\exp (\frac{-\Delta V}{D})  \tag{7}
\end{equation}

where $\gamma _{0}$ is a prefactor, $\Delta $V and D are the height of a
potential barrier and a noise intensity, respectively. Obviously, $\tau
_{K}=1/\gamma _{K}$ and therefore $\tau _{K}$ is adjustable through the
noise intensity D. The match between $\tau _{\omega }$ and $\tau _{K}$
brings about the common SR [9], whereas the match between $\tau _{s}$ and $%
\tau _{K}$ causes a noise-induced nonequilibrium phase transition. It is the
match of three competing time scales, $\tau _{s}$, $\tau _{\omega }$ and $%
\tau _{K}$, that gives rise to the nonequilibrium dynamic phase transition
which we address in this paper. For the sake of simplicity, we assume a
fixed amplitude and frequency of the driving field and a constant reduced
temperature. Hence, $\tau _{\omega }$ and $\tau _{s}$ are invariable in
discussion below. Set $\omega =2$ then $\tau _{\omega }=\pi $. According to
either the theory of para- and ferro-magnetic resonance [22] or the linear
responding theory [23], one can work out an analogic temperature dependence
of $\tau _{s}$ out of either theory above. The $\tau _{s}$ we figure out by
means of the linear responding theory [23] is between 2.5\symbol{126}6
within a wide range of temperature. We plot in Fig. 4 the variation of $\tau
_{K}$ versus the additive noise intensity A using a reduced equation 7, $%
\tau _{K}=1/(\gamma _{K}/\gamma _{0})$. For the TDGL model given by eq.
(2-2), $\Delta V=1/4$. It is easily observed from Fig. 4 that if $A<0.25$,
the $\tau _{K}$ is nearly divergent, and when $0.25<A<1$, $\tau _{K}$ is
between 1.5\symbol{126}10 which matches with $\tau _{\omega }$ and $\tau
_{s} $ in an order of magnitude. When $A>1$, $\tau _{K}=1$. The trend of $%
\tau _{K}$ against the additive noise intensity A in Fig. 4 complies
excellently with the Q\symbol{126}A trend demonstrated in Fig. 2. The
optimal additive noise A$_{\text{P}}$ at which the dynamic order parameter Q
has a peak value is around 0.25\symbol{126}0.5. The effect of a
multiplicative noise, similar to the trend of an additive noise as mentioned
above, is more noticeable in that the hopping of a Brownian particle
agitated by a multiplicative noise is even modulated by the instant order
parameter m(t) itself. Because of $|m(t)|\leqslant 1$ as well as the m(t)
declines with the rise of temperature, the multiplicative noise must satisfy
the requirement of $M_{P}>A_{P}$ so as to achieve the same effect that an
additive noise has. Fig. 3 shows that the optimal multiplicative noise M$_{%
\text{P }}$is about 3, greater than its counterpart A$_{\text{P}}$.
Considering that the modulation the order parameter m impacts on the
multiplicative noise intensity M, one can understand why the dynamic order
parameter Q relies upon the M more drastically than the A when $M>Mp$. And
that is the reason why the range of reentrance induced by a multiplicative
noise is much narrower than that caused by an additive noise, as typically
shown in Figs. 2 and 3. The noise-induced dynamic ordering and the NDPT come
into being only when $\tau _{K}$ complies someway with $\tau _{\omega }$and $%
\tau _{s}$ in quantity. In a word, the matches among the three competing
time scales, what we deal with and present in this paper, are the
fundamental condition for the occurrence of a dynamic ordering, and hence
the nonequilibrium dynamic phase transition in a kinetic Ising spin system
subject to a deterministic external field and a correlated noise source
simultaneously.

\bigskip

\subsection{\protect\bigskip \textbf{Conclusions}}

The kinetic Ising spin system, when driven by a deterministic external field
and a stochastic source of correlated additive and multiplicative noise
simultaneously, holds a characteristic of stochastic resonance and gives
rise to the nonequilibrium dynamic phase transition. Within a
multi-parameter space consisting of the amplitude of a deterministic
external field, the intensities of additive and multiplicative noises, the
relevant correlation intensity between the two sorts of noises together, the
dynamic order parameter take an apparent reentrant trend against noise
intensity, indicating the development and the fading away of dynamic
ordering due to the variation from a cooperative match to a mismatch among
the deterministic driving field, stochastic source and the intrinsic
relaxation of a spin system itself. A multiplicative noise has a more
obvious effect on inducing the dynamic order in a kinetic Ising spin system
than an additive noise does.

\subsection{Acknowledgements}

This work was supported by the National Natural Science Foundation of China
(Grant No. 60471023) and the Natural Science Foundation of Guangdong
Province (Grant No. 031544)

\bigskip

\bigskip \bigskip \bigskip

\bigskip

\bigskip

\bigskip

\bigskip

\bigskip

\bigskip

\bigskip

\bigskip

\bigskip

\bigskip

\bigskip

\bigskip

\bigskip

\bigskip

\bigskip

\bigskip

\bigskip

\bigskip

\bigskip

\subsection{\protect\bigskip Figures}

\bigskip

\bigskip \bigskip

\bigskip

\FRAME{ftbpFU}{2.6705in}{2.1482in}{0pt}{\Qcb{figure1: (a) A=M=0.5, h0=0.0, $%
\protect\lambda $=0.0}}{}{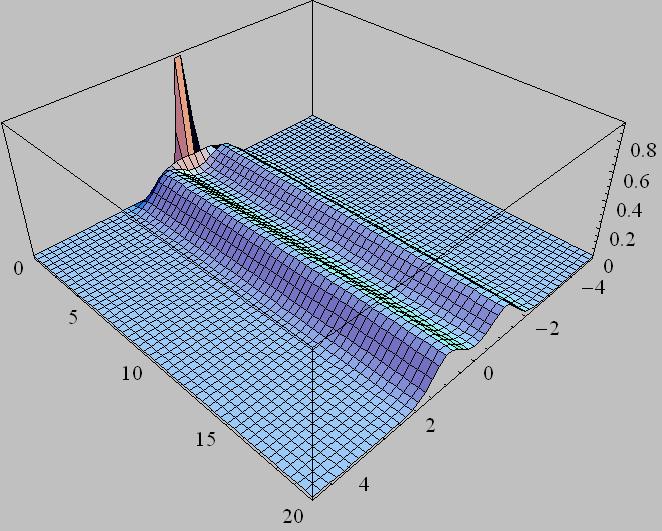}{\special{language "Scientific
Word";type "GRAPHIC";maintain-aspect-ratio TRUE;display "USEDEF";valid_file
"F";width 2.6705in;height 2.1482in;depth 0pt;original-width
2.207in;original-height 1.7703in;cropleft "0";croptop "1";cropright
"1";cropbottom "0";filename 'fig1a.jpg';file-properties "XNPEU";}}

\bigskip \FRAME{ftbpFU}{2.6913in}{2.1638in}{0pt}{\Qcb{figure1: (b) A=M=0.5,
h0=0.2, $\protect\lambda $=0.0}}{}{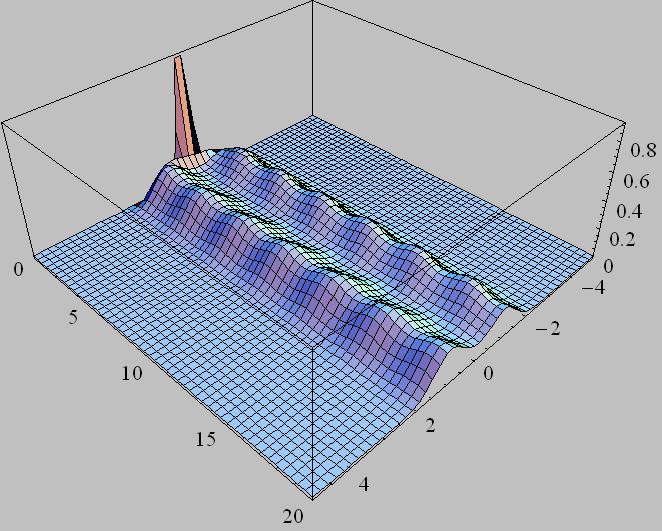}{\special{language "Scientific
Word";type "GRAPHIC";maintain-aspect-ratio TRUE;display "USEDEF";valid_file
"F";width 2.6913in;height 2.1638in;depth 0pt;original-width
2.207in;original-height 1.7703in;cropleft "0";croptop "1";cropright
"1";cropbottom "0";filename 'fig1b.jpg';file-properties "XNPEU";}}

\bigskip \FRAME{ftbpFU}{2.7397in}{2.2044in}{0pt}{\Qcb{figure1: (c) A=M=0.5,
h0=0.2, $\protect\lambda $=0.5}}{}{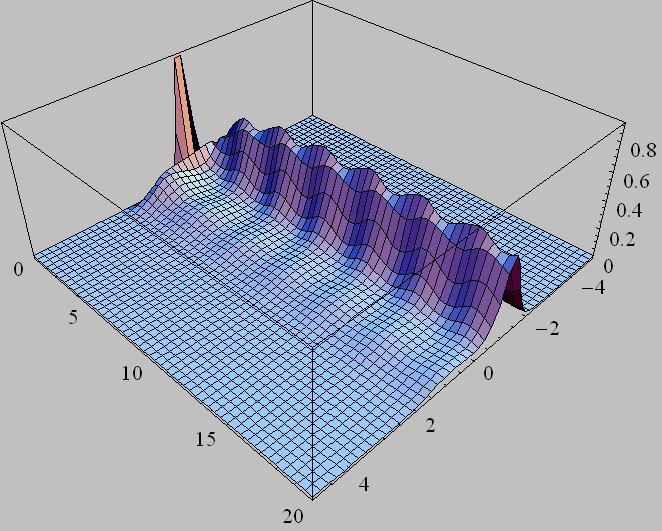}{\special{language "Scientific
Word";type "GRAPHIC";maintain-aspect-ratio TRUE;display "USEDEF";valid_file
"F";width 2.7397in;height 2.2044in;depth 0pt;original-width
2.207in;original-height 1.7703in;cropleft "0";croptop "1";cropright
"1";cropbottom "0";filename 'fig1c.jpg';file-properties "XNPEU";}}

\bigskip \FRAME{ftbpFU}{2.5875in}{1.8325in}{0pt}{\Qcb{Fig. 1.The
distribution function of probability p(m,t) versus instant order parameter m
and evolution time t at zero driving field and zero correlation of noise
(a), nonzero driving field and zero correlation of noise (b), and nonzero
driving field and nonzero correlation of noise (c).}}{}{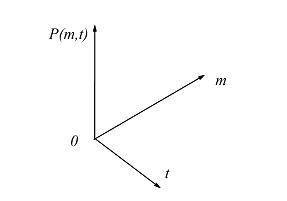}{\special%
{language "Scientific Word";type "GRAPHIC";maintain-aspect-ratio
TRUE;display "USEDEF";valid_file "F";width 2.5875in;height 1.8325in;depth
0pt;original-width 0.9565in;original-height 0.6702in;cropleft "0";croptop
"1";cropright "1";cropbottom "0";filename 'fig1.jpg';file-properties
"XNPEU";}}

\bigskip \FRAME{ftbpFU}{3.9314in}{3.3572in}{0pt}{\Qcb{figure2: (a)}}{}{%
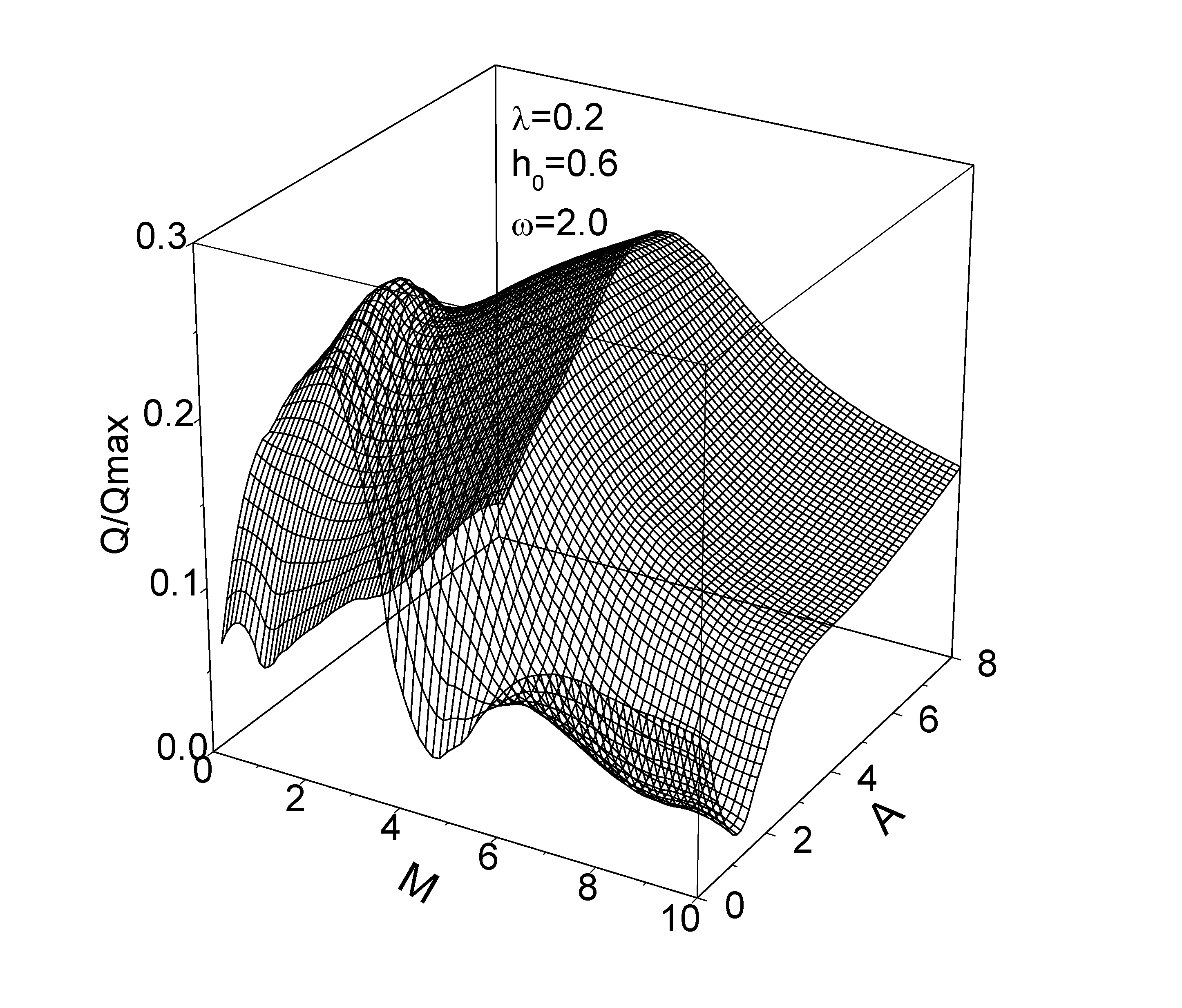}{\special{language "Scientific Word";type
"GRAPHIC";maintain-aspect-ratio TRUE;display "USEDEF";valid_file "F";width
3.9314in;height 3.3572in;depth 0pt;original-width 9.6435in;original-height
8.227in;cropleft "0";croptop "1";cropright "1";cropbottom "0";filename
'fig2a.png';file-properties "XNPEU";}}\FRAME{ftbpFU}{3.9479in}{3.147in}{0pt}{%
\Qcb{figure2: (b)}}{}{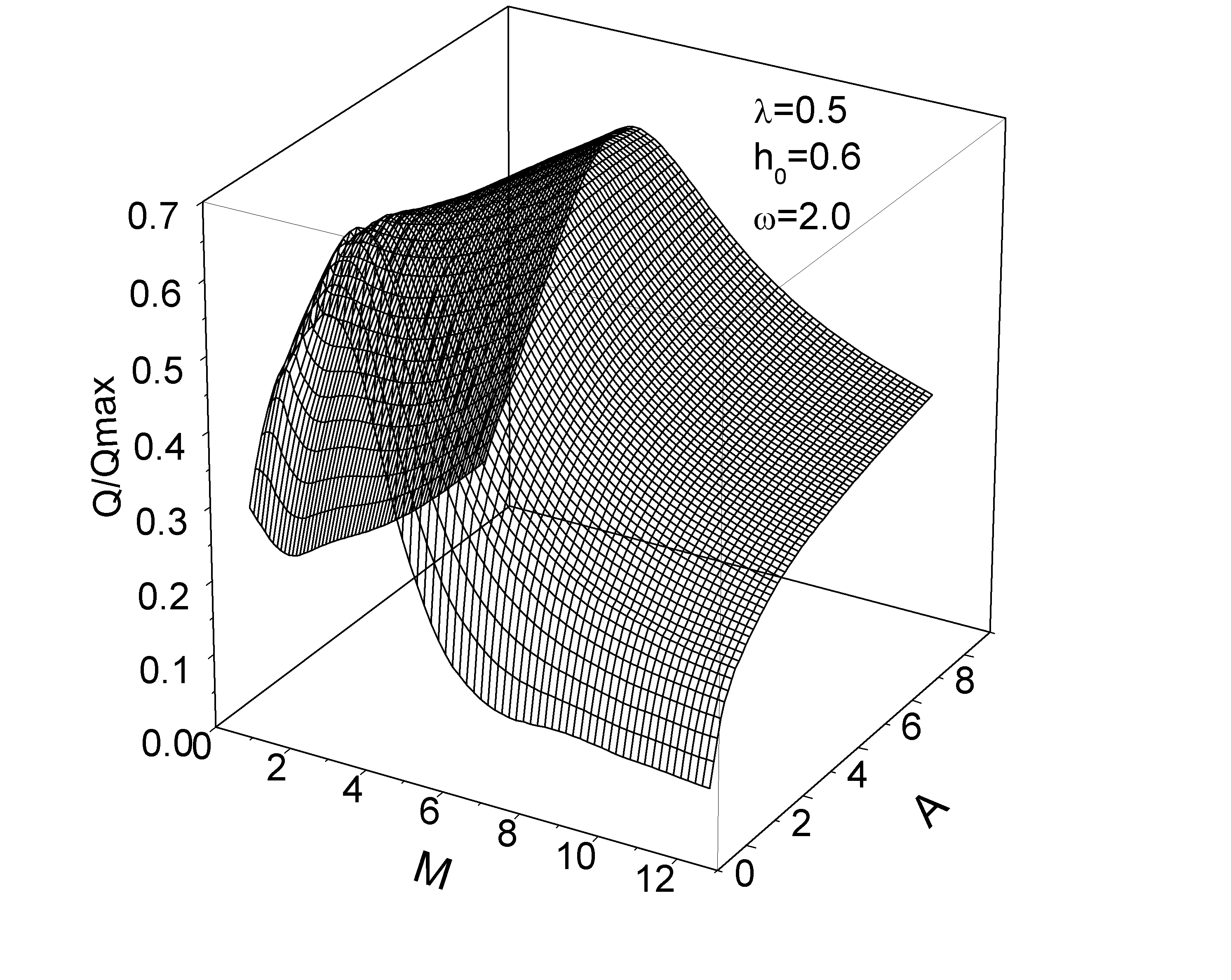}{\special{language "Scientific Word";type
"GRAPHIC";maintain-aspect-ratio TRUE;display "USEDEF";valid_file "F";width
3.9479in;height 3.147in;depth 0pt;original-width 9.9531in;original-height
7.9234in;cropleft "0";croptop "1";cropright "1";cropbottom "0";filename
'fig2b.png';file-properties "XNPEU";}}

\bigskip

\bigskip \FRAME{ftbpFU}{4.1407in}{3.2456in}{0pt}{\Qcb{figure 2: (c)}}{}{%
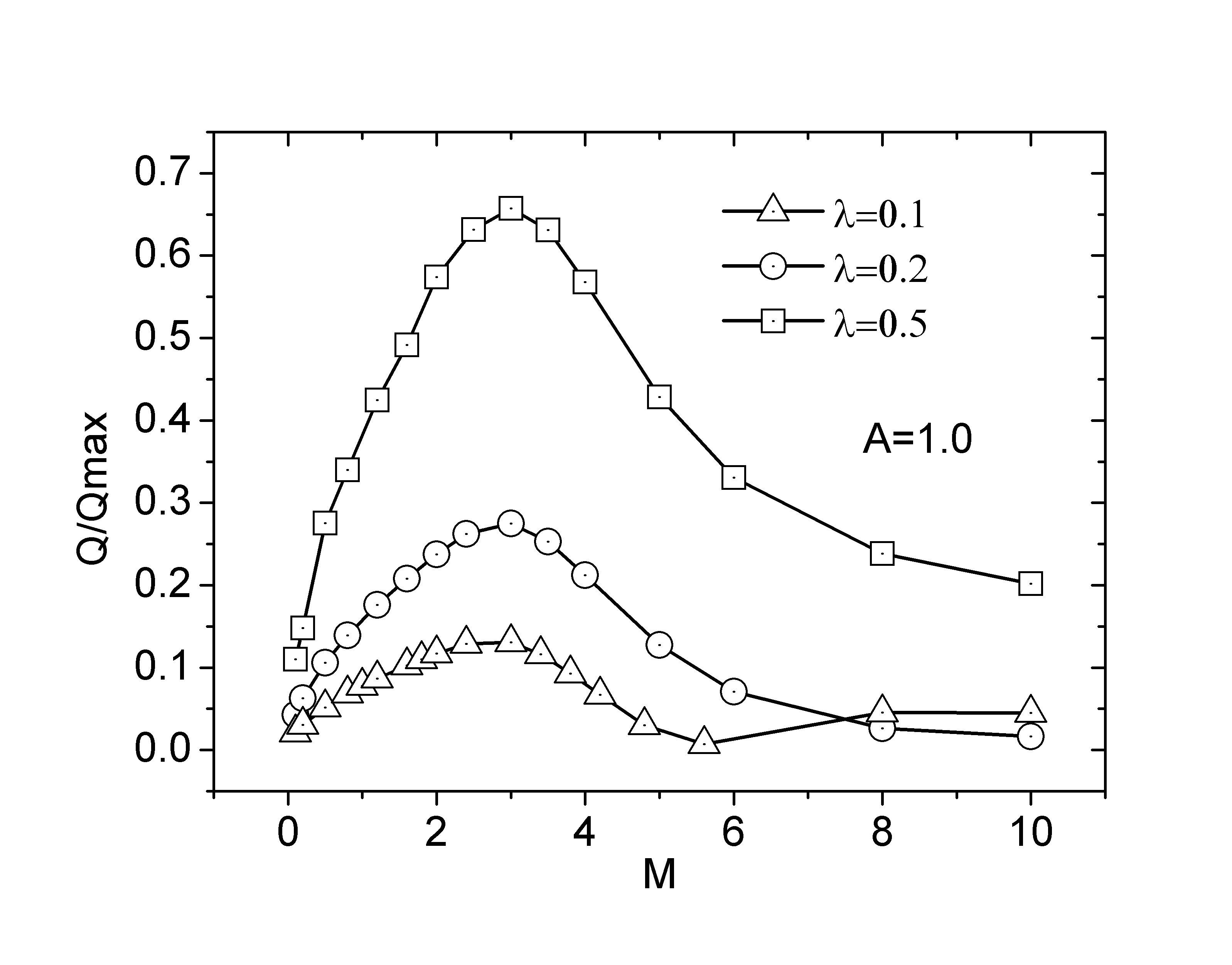}{\special{language "Scientific Word";type
"GRAPHIC";maintain-aspect-ratio TRUE;display "USEDEF";valid_file "F";width
4.1407in;height 3.2456in;depth 0pt;original-width 10.6502in;original-height
8.3403in;cropleft "0";croptop "1";cropright "1";cropbottom "0";filename
'fig2c.png';file-properties "XNPEU";}}

\bigskip \FRAME{ftbpFU}{4.1061in}{3.2024in}{0pt}{\Qcb{figure 2: (d) \ \ \ \
\ \ \ \ \ \ \ \ \ \ \ \ \ \ \ Fig2. The dependence of dynamic order Q upon
additive noise intensity A and multiplicative noise M at different
correlation intensities of noises ($\protect\lambda $=0.2, $\protect\lambda $%
=0.5).}}{}{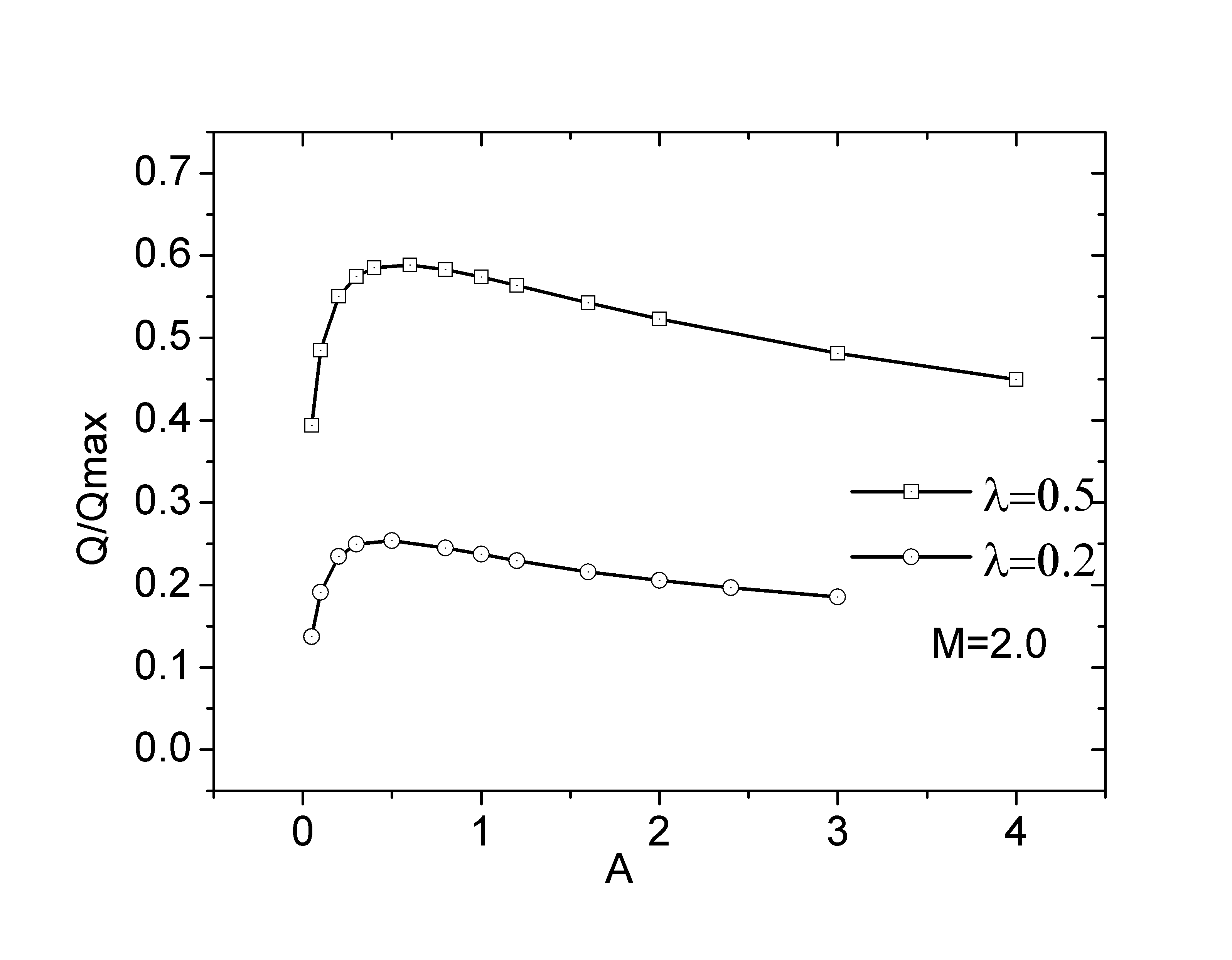}{\special{language "Scientific Word";type
"GRAPHIC";maintain-aspect-ratio TRUE;display "USEDEF";valid_file "F";width
4.1061in;height 3.2024in;depth 0pt;original-width 10.6502in;original-height
8.3005in;cropleft "0";croptop "1";cropright "1";cropbottom "0";filename
'fig2d.png';file-properties "XNPEU";}}

\bigskip \FRAME{ftbpFU}{4.0093in}{3.1047in}{0pt}{\Qcb{figure 3: (a)}}{}{%
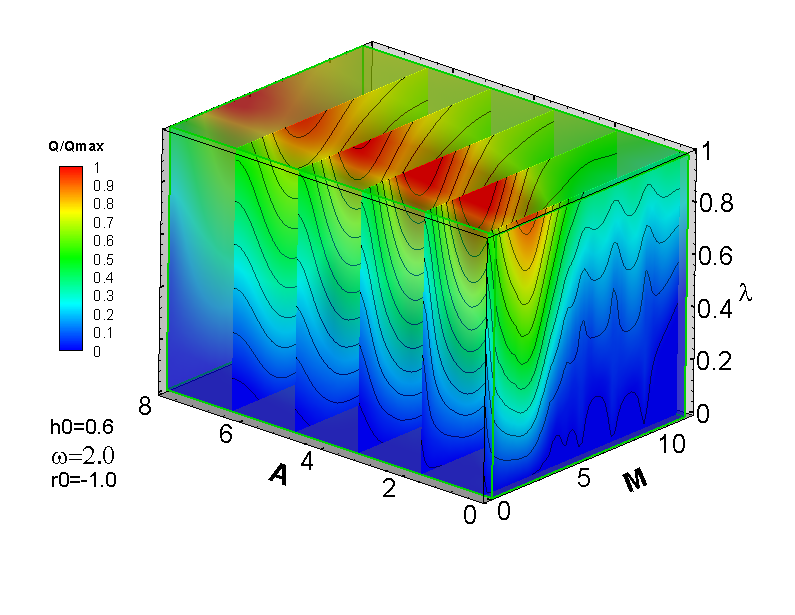}{\special{language "Scientific Word";type
"GRAPHIC";maintain-aspect-ratio TRUE;display "USEDEF";valid_file "F";width
4.0093in;height 3.1047in;depth 0pt;original-width 5.3004in;original-height
4.1001in;cropleft "0";croptop "1";cropright "1";cropbottom "0";filename
'fig3a.png';file-properties "XNPEU";}}

\bigskip \FRAME{ftbpFU}{3.9972in}{3.0995in}{0pt}{\Qcb{figure 3: (b) \ \ \ \
\ \ \ \ \ \ \ \ \ \ \ \ \ \ Fig. 3. The sliced contour of dynamic order
parameter Q within the parameter space of A, M and $\protect\lambda $
(above), of A, M, and h0 (below)}}{}{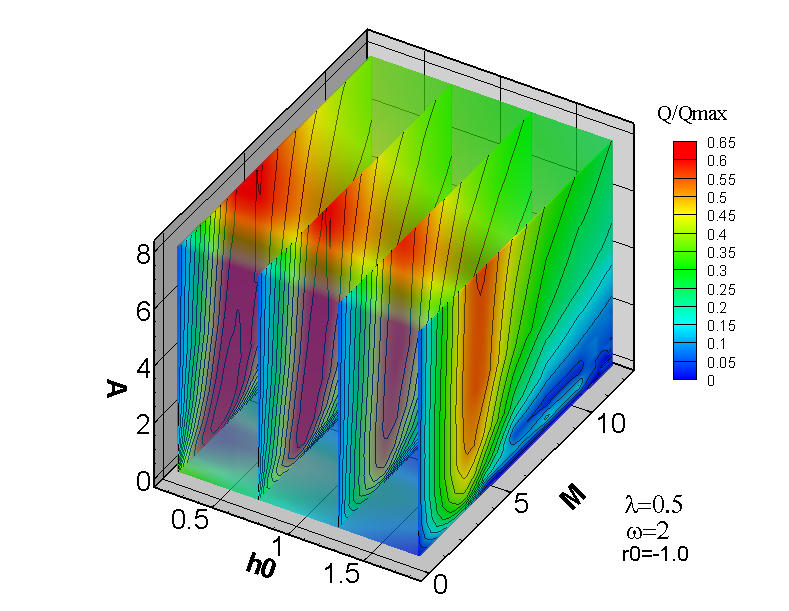}{\special{language
"Scientific Word";type "GRAPHIC";maintain-aspect-ratio TRUE;display
"USEDEF";valid_file "F";width 3.9972in;height 3.0995in;depth
0pt;original-width 5.2607in;original-height 4.0733in;cropleft "0";croptop
"1";cropright "1";cropbottom "0";filename 'fig3b.png';file-properties
"XNPEU";}}

\bigskip \FRAME{ftbpFU}{3.7308in}{2.8461in}{0pt}{\Qcb{Figure 4. Reduced $%
\protect\tau _{\text{K}}$ versus additive noise intensity A.}}{}{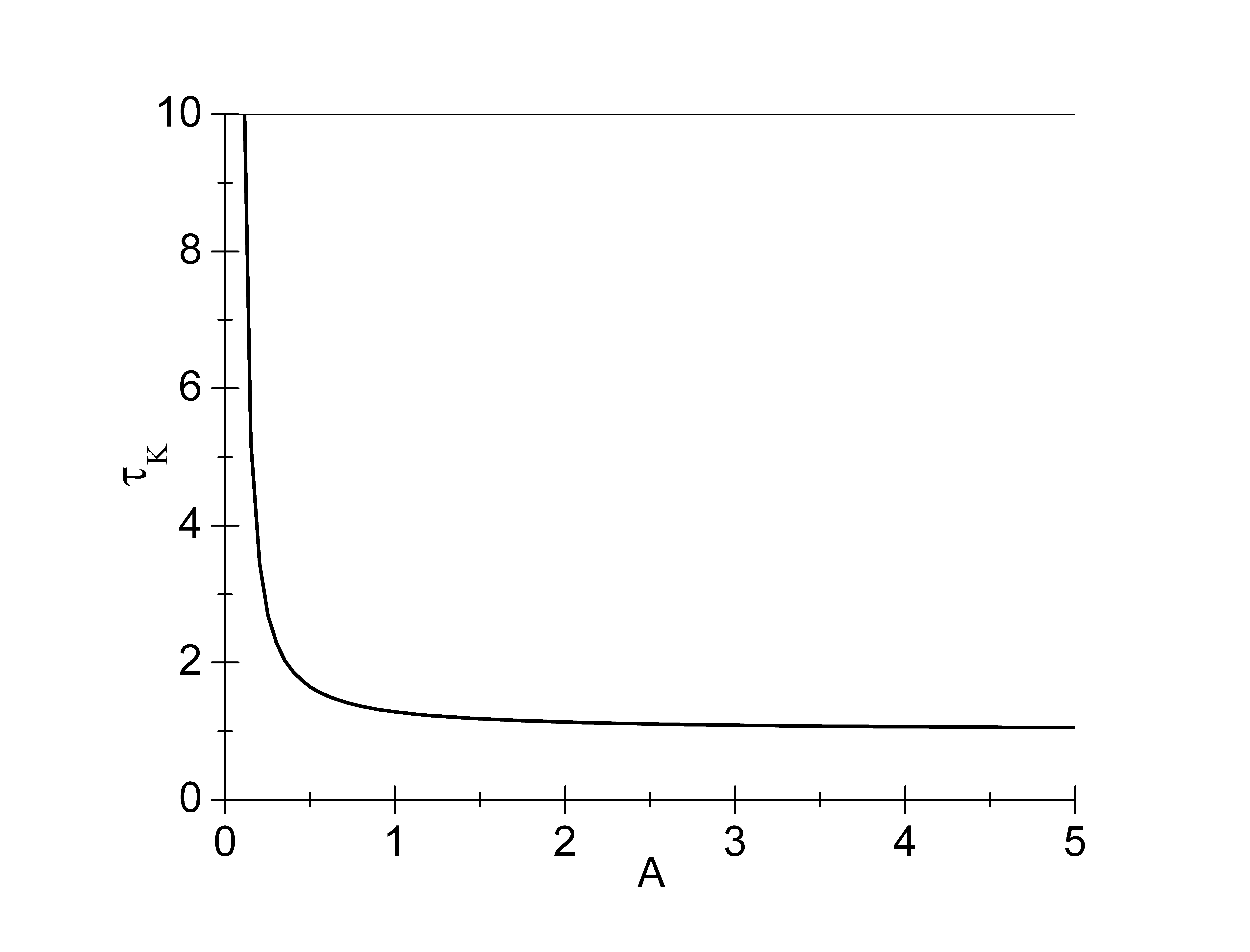}{%
\special{language "Scientific Word";type "GRAPHIC";maintain-aspect-ratio
TRUE;display "USEDEF";valid_file "F";width 3.7308in;height 2.8461in;depth
0pt;original-width 41.9996in;original-height 31.9998in;cropleft "0";croptop
"1";cropright "1";cropbottom "0";filename 'fig4.png';file-properties
"XNPEU";}}

\bigskip

\end{document}